\documentclass[reprint,superscriptaddress, aps,longbibliography]{revtex4-1}

\usepackage{blindtext}
    \usepackage{amsmath}
    \usepackage{makeidx}
    \usepackage{amsfonts}
    \usepackage{float}
    \usepackage[usenames,dvipsnames]{pstricks}
    \usepackage{subfigure}
    \usepackage{epsfig}
    \usepackage{pst-grad} 
    \usepackage{pst-plot} 
    \usepackage[colorlinks,hyperindex]{hyperref}
    \hypersetup
    {
        colorlinks,%
        citecolor=black,%
        linkcolor=black,%
        urlcolor=black,%
    }

\usepackage{color}
\usepackage{dsfont}
\usepackage{graphicx}
\usepackage{subfigure}

\usepackage{dcolumn}
\usepackage{amsmath}    
\usepackage{verbatim}   
\usepackage{color}      
\usepackage{hyperref}   
\usepackage{amsfonts}
\usepackage{braket}
\usepackage{gensymb}
\usepackage{bm}
\usepackage{epstopdf}
\usepackage{titlesec}
\usepackage{capt-of}
\usepackage{float}

\begin{document}


\title{
Photon phase shift at the few-photon level and optical switching by a quantum dot in a microcavity
}

\author{L. M. Wells}\,
\email{lw585@cam.ac.uk}
\affiliation{Toshiba Research Europe Limited, Cambridge Research Laboratory,\\
208 Science Park, Milton Road, Cambridge, CB4 0GZ, U.K.}
\affiliation{Cavendish Laboratory, Cambridge University,\\
J. J. Thomson Avenue, Cambridge, CB3 0HE, U.K.}

\author{S. Kalliakos}
\affiliation{Toshiba Research Europe Limited, Cambridge Research Laboratory,\\
208 Science Park, Milton Road, Cambridge, CB4 0GZ, U.K.}

\author{B. Villa}\,
\affiliation{Toshiba Research Europe Limited, Cambridge Research Laboratory,\\
208 Science Park, Milton Road, Cambridge, CB4 0GZ, U.K.}
\affiliation{Cavendish Laboratory, Cambridge University,\\
J. J. Thomson Avenue, Cambridge, CB3 0HE, U.K.}

\author{D. J. P. Ellis}
\affiliation{Toshiba Research Europe Limited, Cambridge Research Laboratory,\\
208 Science Park, Milton Road, Cambridge, CB4 0GZ, U.K.}

\author{R. M. Stevenson}
\affiliation{Toshiba Research Europe Limited, Cambridge Research Laboratory,\\
208 Science Park, Milton Road, Cambridge, CB4 0GZ, U.K.}

\author{A. J. Bennett}
\thanks{Current affiliation: Institute for Compound Semiconductors, Cardiff University, Queen's Buildings, 5 The Parade, Roath, Cardiff, CF24 3AA, U.K.}
\affiliation{Toshiba Research Europe Limited, Cambridge Research Laboratory,\\
208 Science Park, Milton Road, Cambridge, CB4 0GZ, U.K.}

\author{I. Farrer}
\thanks{Current affiliation: Department of Electronic \& Electrical Engineering, University of Sheffield, Mappin Street, Sheffield, S1 3JD, U.K. }
\affiliation{Cavendish Laboratory, Cambridge University,\\
J. J. Thomson Avenue, Cambridge, CB3 0HE, U.K.}

\author{D. A. Ritchie}
\affiliation{Cavendish Laboratory, Cambridge University,\\
J. J. Thomson Avenue, Cambridge, CB3 0HE, U.K.}

\author{A. J. Shields}
\affiliation{Toshiba Research Europe Limited, Cambridge Research Laboratory,\\
208 Science Park, Milton Road, Cambridge, CB4 0GZ, U.K.}

\date{\today}%

\begin{abstract}
We exploit the nonlinearity arising from the spin-photon interaction in an InAs quantum dot to demonstrate phase shifts of scattered light pulses at the single-photon level. Photon phase shifts of close to 90$\degree$ are achieved using a charged quantum dot in a micropillar cavity. We also demonstrate a photon phase switch by using a spin-pumping mechanism through Raman transitions in an in-plane magnetic field. The experimental findings are supported by a theoretical model which explores the dynamics of the system. Our results demonstrate the potential of quantum dot-induced nonlinearities for quantum information processing.      
\end{abstract}

\maketitle

\section{Introduction}

Semiconductor quantum dots (QDs) are considered a promising candidate for quantum information processing. As excellent sources of single photons, they demonstrate unparalleled brightness \cite{Maier:14}, near perfect indistinguishability \cite{Somaschi.2016,PhysRevLett.116.020401} and excellent efficiency \cite{Claudon.2010}. They can be embedded in a variety of nanophotonic structures for enhanced light-matter interaction \cite{Gerard.1996,Kiraz.2001,Vuckovic.2003}. The non-linear effects arising from the interaction between a photon and a single charge spin confined in a QD can be used to achieve a range of quantum operations required for quantum information processing. To that aim, spin-photon entanglement has been recently demonstrated \cite{Togan.2010}, while other applications such as logic operations \cite{PhysRevLett.104.160503,PhysRevLett.114.173603,Bennett.2016,Hu.2017} have been proposed.  

There has been a significant effort to exploit the nonlinearities arising from spin-photon interactions to realize a quantum switch. Proposals have been made to make a spin-photon switch using an emitter in a cavity \cite{PhysRevLett.92.127902,PhysRevLett.93.250502}. Typically, this relies on the rotation of the polarization of a photon coherently scattered by the single spin, inducing photon phase shifts $\phi$ up to 180$\degree$. This so-called giant Faraday or Kerr rotation began to attract attention in the 1980s, when theoretical proposals suggested utilising the phenomenon to achieve optical quantum non-demolition measurements \cite{Grangier.1989,Grangier.1998,PhysRevA.60.4974}. Measurements of Kerr and Faraday rotations using QDs were not reported until much later, with rotation angles in the few $10^{-3}$ degree range \cite{Berezovsky.2006,Atature.2007,Mikkelsen.2007} reported on charged QD systems under a Faraday geometry magnetic field. Significant improvement on the rotation angles were reported recently, with rotations of 6$\degree$ recorded for a QD strongly coupled to a micropillar cavity at T = 20 K \cite{Arnold.2015}. More recently, rotations of more than 90$\degree$ were reported for a QD in a ``bad" cavity in a Faraday geometry magnetic field \cite{Androvitsaneas:15}.  A quantum phase switch using a QD was demonstrated for the first time in 2016, using a QD strongly coupled to a photonic crystal defect cavity in a Voigt geometry magnetic field \cite{Sun.2016}. While 2D photonic crystal cavities can offer high Q-factors and integration with on-chip quantum photonic circuits, they have low photon extraction efficiencies compared to micropillar structures \cite{Liu.2017}. Furthermore, achieving strong light-matter coupling is demanding, with limited reproducibility considering current nanofabrication processes.  

Here, we demonstrate a photon phase switch using a charged QD weakly coupled to the confined mode of a micropillar cavity. Using a Voigt geometry magnetic field, we demonstrate phase shifts of $80\pm 2\degree$ of coherently scattered laser pulses which, on average, contain $\sim$ one photon. After preparing the spin in an eigenstate \cite{Atature.2006}, we use a second pulse to switch the QD-induced phase shift of the initial pulse ON and OFF. Finally, we develop a theoretical model that provides further insight into performance limiting factors. We show that polarization control is possible using a transition in a single quantum dot \cite{PhysRevX.8.021078}.   

\section{Concept and Experimental Setup}

\begin{figure*}[t]
\includegraphics[width=\textwidth]{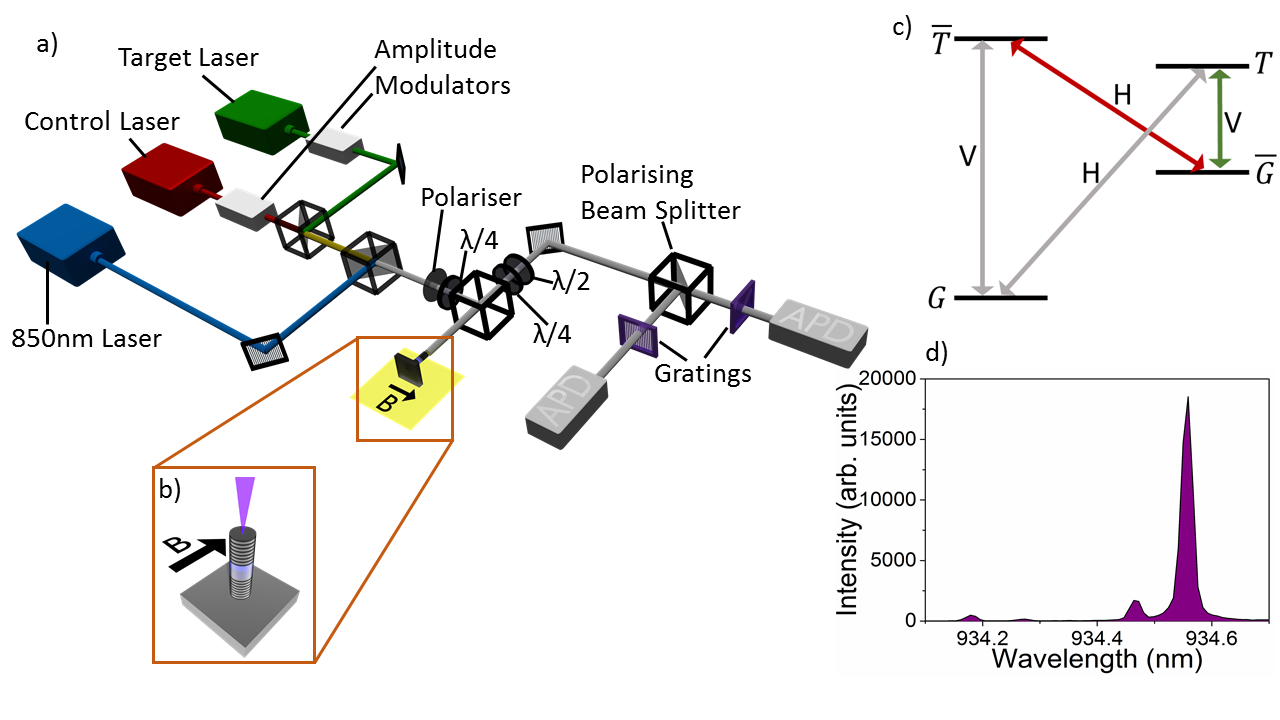}
\caption{ 	 
{\bf a)} Illustration of our experimental setup. {\bf b)} Illustration of a micropillar cavity in a Voigt geometry magnetic field. {\bf c)} Energy level diagram, showing the cavity-enhanced``Target" transition (green arrow) and ``Control" transition (red arrow). $\bf d)$ Photoluminescence spectrum of the QD under non-resonant excitation. The long wavelength transition is enhanced by the cavity mode.}  
\label{Fig1}
\end{figure*}

The core of our phase-switching system is a singly-charged InAs quantum dot held in a B = 8 T magnetic field (Voigt geometry) \cite{doi:10.1063/1.4719077}. Its energy levels form a double lambda system (Figure \ref{Fig1}c). $\ket{G}$ and $\ket{\bar{G}}$ ($\ket{T}$ and $\ket{\bar{T}}$) represent orthogonal ground (trion) states. The QD is in a 2.75 $\mu$m diameter micropillar cavity with the top (bottom) mirror consisting of 17 (25) mirror pairs. This increases the light collection efficiency and enhances emission from the long-wavelength vertical transition of interest through weak coupling with the cavity mode, with Purcell factor $F_P \approx 2$ and cooperativity $C=\frac{1}{2}F_P \approx 1$ \cite{Santis.2017}. The cavity mode quality factor is $Q \approx 5000$. The vertical and diagonal transitions couple to orthogonal linear polarizations of light, represented by V and H respectively. The long-wavelength vertical transition is excited using resonant right-handed circularly polarized laser pulses. The quantum state of the photon can be written as $\ket{\phi_i}= \ket{H}+i\ket{V}$ \cite{loudon2000quantum}.  On reflection, the state becomes $\ket{\phi_f}_{\uparrow(\downarrow)}=\ket{H}+ir_{\uparrow(\downarrow)} \ket{V}$, where $r_{\uparrow}$ ($r_\downarrow$) is the reflection coefficient for the QD in the spin-up (spin-down) state. The observed phase change is dependent on the interference contrast $\alpha = \frac{\kappa_{ex}}{\kappa}$, where $\kappa_{ex}$ and $\kappa$ are the cavity energy decay rate to the reflected mode and the total cavity energy decay rate. When $\alpha>0.5$ and $C>2\alpha-1$, $r_{\uparrow}$ and $r_{\downarrow}$ have opposite signs. Hence the incident photon experiences a spin state conditional 180$\degree$ phase shift \cite{michler2017quantum}. Fitting the reflectivity spectrum of our cavity allows us to extract a value of $\alpha = 0.93$ \cite{Sun.2016}.  We define polarization contrast as P = $\frac{I_R-I_L}{I_R+I_L}$, where $I_R$ and $I_L$ are the scattered intensities of right- (RCP) and left- (LCP) circularly polarized light respectively. The QD-induced photon phase shift is determined by $\phi=cos^{-1}(P)$. 

Our experimental setup is illustrated in Figure 1a.  The enhanced vertical $\ket{\bar{G}}\rightarrow\ket{T}$ transition is probed using a narrow-linewidth laser at 934.55 nm, referred to as the ``Target".  A second narrow-linewidth laser at 934.49nm can be used to drive the diagonal $\ket{\bar{G}}\rightarrow\ket{\bar{T}}$ transition, and termed the ``Control" (Figure \ref{Fig1}c).  Emission from these lasers is controlled via amplitude modulators to achieve coherent pulses (250 ps (7ns) width for the ``Target" (``Control")) with 80 MHz repetition rate. The excitation pulses are RCP.  A series of polarization optics and a polarizing beam splitter in the detection path allow us to record $I_R$ and $I_L$ simultaneously to measure the individual contribution of each polarization to the total collected light. As the sample is nominally undoped, we use a weak non-resonant pulsed laser to inject a charge with random spin into the QD. \cite{PhysRevX.8.021078}. Both positive and negative trions are created in this way, although only one is resonant with the cavity and observed here (see Suplemental Material for further detail \cite{supmat}). The photoluminescence spectrum under non-resonant excitation at B = 8 T is shown in Figure \ref{Fig1}d. We observe four distinct peaks that are assigned to the vertical and diagonal transitions, with the longest wavelength transition enhanced due to weak coupling with the cavity mode. The inferred g-factor is $\lvert 0.86 \rvert$ ($\lvert 0.18 \rvert$) for the ground states (excited states). These values are comparable with previous reports detailing both electrons and holes in the ground state \cite{Bennett.gfactor,PhysRevLett.98.047401,PhysRevB.93.035311}. It is not possible to assign a specific carrier to the g-factors from Figure \ref{Fig1}d and our analysis. Since the injected charge type does not impact the mechanisms used for this experiment, we do not specify whether an electron or hole is captured. Temperature is used to tune the long-wavelength vertical transition in resonance with the cavity mode. The sample is therefore held at T = 18 K.

\section{Quantum dot-induced photon phase shift}

\begin{figure}[t!]
\begin{minipage}[h]{0.5\textwidth}
\centering
\includegraphics[width=\textwidth]{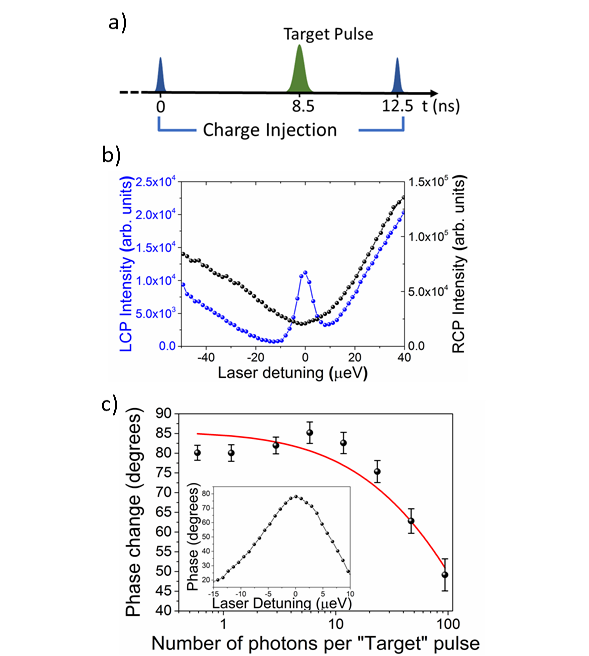}
\end{minipage}
\caption{{\bf a)} The pulse sequence used to phase shift the reflected photons. {\bf b)} The reflectivity spectrum of the cavity for LCP (blue) and RCP (black) light, with average 12 photons per ``Target" pulse. {\bf c)} Recorded phase shift as a function of the average number of photons per ``Target" pulse (black data points) and the calculated phase shift (red line). Inset: Phase change as a function of laser detuning for an average of 12 photons per ``Target" pulse.}
\label{Fig2}  
\end{figure}

After injecting a charge into the QD with the weak non-resonant laser pulse, we probe the change in photon phase induced by the cavity coupled QD transition (Figure \ref{Fig2}a). The system is probed using the ``Target" pulse and we measure the RCP and LCP components of the reflected signal (Figure \ref{Fig2}b). The RCP reflected light shows the cavity response and does not contain any indication the incident photon interacted with the QD spin. In the LCP channel, we observe a modification of the cavity reflectivity when the laser is on resonance with the QD transition. The observed peak is due to resonance fluorescence and is a manifestation of the QD phase shift. The marginal split ($\sim$ 6 $\mu$eV) observed between the cavity centers recorded in the two polarizations is attributed to the small ellipticity of the micropillar acquired during the fabrication process. We note that the observed peak measures 10 $\mu$eV in width, and is broad compared to other publications. Several factors may contribute to this, including the width of the ``Target" pulse ($\sim$ 4 $\mu$eV), the Purcell effect reducing the lifetime of the QD \cite{Bennett.2016} and spectral jittering.

The QD-induced phase shift as a function of laser detuning is shown in the inset of Figure \ref{Fig2}c. It is extracted using the reflected RCP and LCP signals in Figure \ref{Fig2}b. Data are fitted with a Gaussian function and the peak of this curve is used to obtain the maximum achieved phase shift. The resulting maximal QD-induced phase shift of the ``Target" pulses as a function of the average photon number per pulse is shown in Figure \ref{Fig2}c. At the limit of a single photon on average per ``Target" pulse, we observe phase shifts of $80 \pm 2\degree$. Given the random nature of charge initialization in our system results in occupation of both ground states with equal probability, the observed photon phase shifts are close to the expected value of 90$\degree$. Intrinsic effects, such as spectral jitter, may limit the induced phase shifts \cite{Liu.2017}. Furthermore, the cavity is birefringent, as can be seen in the different energies of the modes, which may introduce a degree of ellipticity in the incident RCP ``Target" pulse \cite{ding2018coherent}. As the average photon number per pulse increases, we observe a decrease of the recorded phase shifts. We attribute this to the collected RCP intensity increasing more rapidly than the collected LCP intensity as the number of photons per pulse increases. We model this by considering the decreasing contribution of the coherently-scattered photons to the total scattered intensity as a function of incident power following a previously developed model \cite{Bennett}. We use the measured value of $T_1=0.5$ ns for the transition lifetime and the coherence time $T_2=1$ ns as a fitting parameter. Such a high value for the coherence time is reasonable given the coherent nature of the photon scattering event \cite{Bennett}. The result of this model is shown in Figure \ref{Fig2}c (red line) and is in good agreement with our experimental results.

\section{Quantum dot-induced photon phase switch}


\begin{figure}[t!]
\begin{minipage}[h]{0.5\textwidth}
\centering
\includegraphics[width=\textwidth]{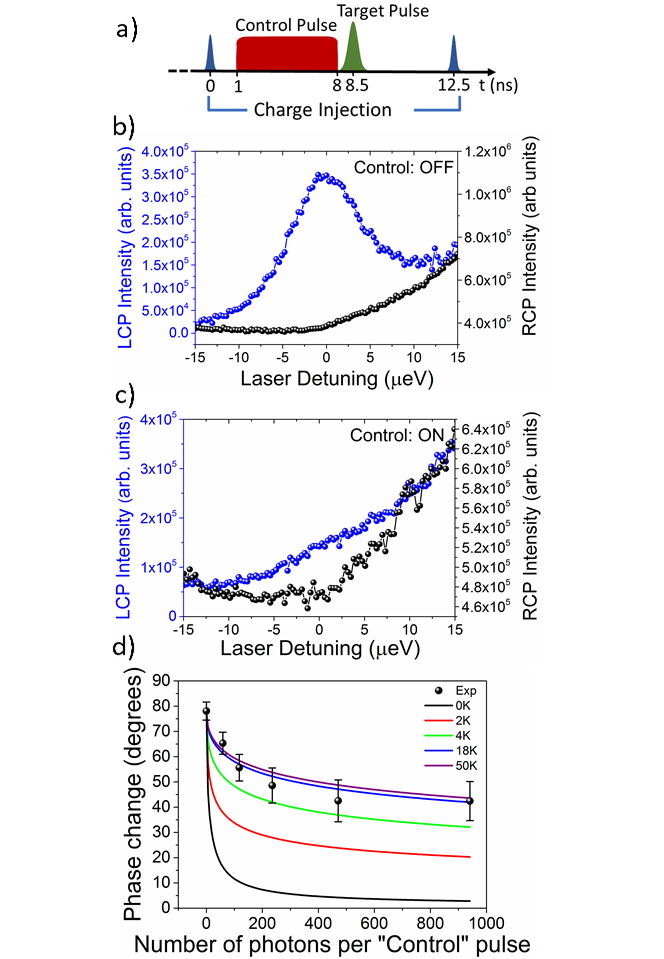}
\end{minipage}
\caption{{\bf a)}  The pulse sequence used to generate and control the phase shift of the reflected photons. {\bf b)} Reflectivity spectrum of the system for RCP (black) and LCP (blue) light when the ``Control" pulse is OFF and {\bf c)} when the ``Control" pulse is ON. Differences in the change in LCP and RCP signals are as a result of a marginal offset between the respective cavity mode centers. {\bf d)} Measured phase change as a function of number of photons per ``Control" pulse (black data points) and calculated phase change for several different temperatures (solid lines).}  
\label{Fig3}
\end{figure}

To demonstrate a QD-induced phase switch, we introduce a ``Control" pulse. The pulse sequence works as follows (Figure \ref{Fig3}a).  The non-resonant weak laser pulse injects a charge into the QD. The ``Control" pulse drives the diagonal $\ket{\bar{G}}\rightarrow\ket{\bar{T}}$ transition, pumping the charge out of the ground state of the $\ket{\bar{G}}\rightarrow\ket{T}$ transition. This pulse duration is relatively long (7 ns) at a relatively high power (average 940 photons per pulse)  to achieve efficient spin pumping.  The ``Target" pulse is used to probe the $\ket{\bar{G}}\rightarrow\ket{T}$ transition, and the phase change is measured. Without a charge in the $\ket{\bar{G}}$ ground state, the ``Target" photon will not interact with the QD and will be reflected without changing phase.  The ``Control" pulse can be switched ON and OFF, enabling control over the observed phase shift. As the two driven transitions are very close ($\sim$ 130 $\mu$eV), extremely narrowband spectral filtering is introduced. This suppresses the contribution of the ``Control" pulse to the recorded signal.  

The spectra for both components of the reflected signal with the ``Control" pulse is OFF (ON) are shown in Figure \ref{Fig3}b (Figure \ref{Fig3}c). In the absence of the control pulse, the LCP reflected signal is enhanced when the driving laser is on resonance with the ``Target" transition, similar to Figure \ref{Fig2}b. Introducing the ``Control" pulse removes this peak almost entirely, effectively acting as a switch and turning OFF the phase rotation. We find the presence of the ``Control" pulse causes a 46\% reduction in the measured phase change, from $78 \pm 4\degree$ to $42 \pm 8\degree$. 

Finally, we vary the average power of the ``Control" laser pulse to investigate its impact on our switching mechanism. In Figure \ref{Fig3}d we plot the phase shift for range of ``Control" pulses with average photon numbers per pulse between 0 and 940. As expected, the phase shift reduces as the number of photons per pulse increases. The induced phase shift reaches a steady value of $42 \pm 8\degree$ once there are 470 or more photons per pulse. This is a far smaller reduction than expected considering a successful spin-pumping mechanism should eliminate the population in the $\ket{\bar{G}}$ state \cite{Atature.2007}.  

To better understand the limiting mechanisms of our photon phase switch, we develop a simple theoretical model based on a system of rate equations for the relevant state populations. \nocite{androvitsaneas2017efficient, PhysRevX.8.021078, Anton:17,PhysRevB.80.201311, Gerardot.2008, Climente.2013} We calculate the polarization contrast (and hence phase shift) using the ratio $\Xi = \frac{N^i_{\bar{G}}-N^f_{\bar{G}}}{N^i_{\bar{G}}+N^f_{\bar{G}}}$, where $N^i_{\bar{G}}$ and $N^f_{\bar{G}}$ are the populations of the $\ket{\bar{G}}$ state before and after the ``Control" pulse respectively (see Supplemental Material for further details on the theoretical model \cite{supmat}). The calculated phase shifts as a function of the average photon number per ``Control" pulse for different temperatures are shown in Fig. \ref{Fig3}d. This suggests that temperature-dependent mechanisms, such as spin-flip and phonon-assisted transitions are the main factors limiting the achieved switching ratio. In particular, we experimentally achieve a phase switching ratio of 46\% for an average of 940 photons per ``Control" pulse (corresponding to $36\degree$ phase shift in Figure \ref{Fig3}d). Without a spin-flip mechanism (T = 0 K), the same ratio is predicted to be 90\% ($76\degree$ phase shift in Figure \ref{Fig3}d) . This is limited by the ellipticity in polarization of the ``Target" pulse, and the non-perfect degeneracy of the cavity modes. For T = 18 K we obtain good agreement between the experimental and theoretical values (blue line Figure \ref{Fig3}c). We anticipate that experiments performed at lower temperatures, or with a higher Zeeman splitting, would reduce the spin-flip rate and therefore improve the switching contrast. We also anticipate that these changes to our system would allow us to significantly reduce the number of photons per ``Control" pulse.  

\section{Conclusions}
We have demonstrated photon phase shifts up to $80 \pm 2\degree$ at the single-photon level by exploiting nonlinear photon-spin interactions in a charged QD in a micropillar cavity. We used Raman transitions allowed due to an external magnetic field to demonstrate controllable switching of the obtained phase shifts. Limitations of the switching mechanism in our system are highlighted by a simple theoretical model based on the rate equations of the relevant states. Our findings highlight the importance of the QD-micropillar cavity system as a nonlinear medium for developing photonic quantum logic operations towards quantum information processing.

\section*{Acknowledgments}
The authors acknowledge funding from the EPSRC for the MBE system used to grow the micropillar cavity. L. W. gratefully acknowledges funding from the EPSRC and financial support from Toshiba Research Europe Ltd. B. V. gratefully acknowledges funding from the European Union's Horizon 2020 research and innovation programme under the Marie Sk\l{}odowska-Curie grant agreement No. 642688 (SAWtrain).

\section*{Data Access}
The experimental data used to produce the figures in this paper is publicly available at https://doi.org/10.17863/CAM.40128.

\renewcommand{\bibname}{References}
\bibliography{references}

\section{Supplementary Information}

\subsection{Quantum dot band gap diagram and charge capture}

\begin{figure}[h!]
\includegraphics[width=7 cm]{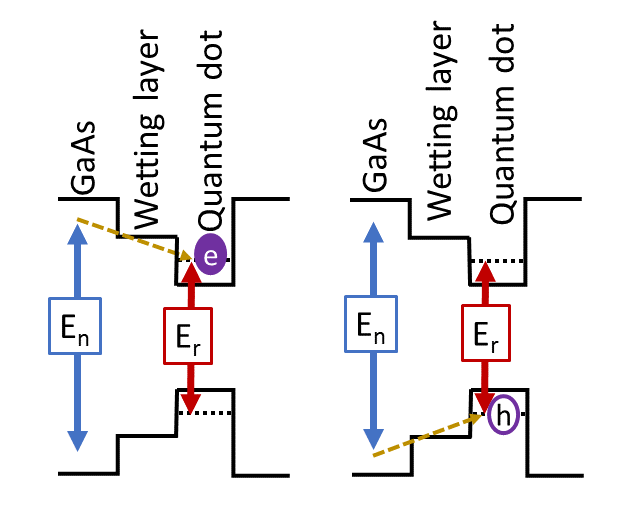}
\caption{Band gap diagram showing electron (left) and hole (right) capture in a QD system.}  
\label{Fig1}
\end{figure}

Band gap diagrams illustrating the process of electron (left) and hole (right) capture are shown in Figure \ref{Fig1}. A weak non-resonant laser E$_n$ creates an electron-hole pair near the QD. One of the charges is then captured by the quantum dot (dashed arrow). The charged quantum dot transitions may then be driven by a resonant laser (the ``Target" or ``Control" in our experiment), shown in Figure \ref{Fig1} as E$_r$. Probabilistically, the non-resonant laser can lead to the creation of both positive and negative trions. However, they will be at different wavelengths and therefore only one trion will be resonant with the cavity mode of our micropillar. Hence, it is only possible for us to see PL from one trion in our experiment. From our data for the electron and hole g-factors and without being able to see additional charge lines, it is not possible for us to determine whether we have a positive or negative trion. For this experiment and others based on similar structures, the observed phenomena do not change based on the type of charge captured by the QD.  

\subsection{Quantum dot-induced photon phase switch theoretical model}

\begin{figure}[h!]
\includegraphics[width=6 cm]{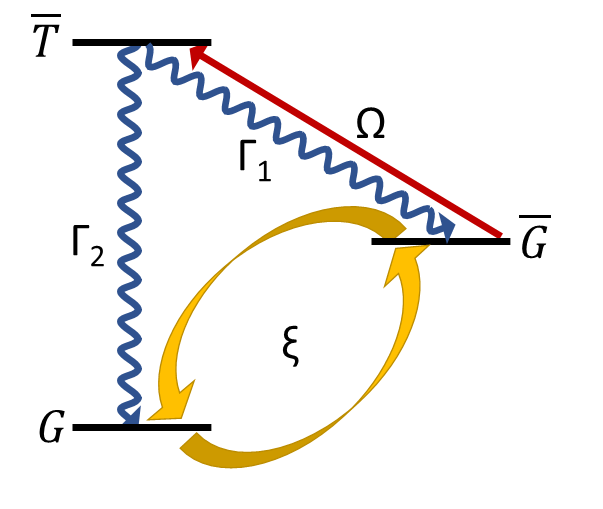}
\caption{A schematic of the pulse sequence used to control the recorded phase shift.}  
\label{Fig3}
\end{figure}

In Figure 3d of the main text, we present the results of a theoretic model used to gain a better understanding of the limitying mechanisms of our system. It is based on a system of rate equations for the population of the relevant states. A schematic of the lambda system used as the basis of our model is shown in Figure \ref{Fig3}. We define $\Omega$ as the Rabi frequency of the "Control" pulse, and $\Gamma_1$ ($\Gamma_2$) as the relaxation rate from the trion state to the $\ket{\bar{G}}$ ($\ket{G}$) ground state. The transition rate between the two ground states is given by $\xi$. We obtain a set of three equations that relate the population of each level of our model to the relevant frequencies and decay rates:
\begin{equation}
\frac{dN_{\bar{G}}}{dt}=\Gamma_1N_{\bar{T}}-\Omega N_{\bar{G}}-\xi N_{\bar{G}}+\xi N_{G}   \end{equation}
\begin{equation}
\frac{dN_G}{dt}=\Gamma_2N_{\bar{T}}-\xi N_G+\xi N_{\bar{G}}    
\end{equation}

\begin{equation}
\frac{dN_{\bar{T}}}{dt}=\Omega N_{\bar{G}}-\Gamma_1 N_{\bar{T}}-\Gamma_2 N_{\bar{T}}   
\end{equation}

where $N_{\bar{G}}$ represents the population of the $\ket{\bar{G}}$ state, $N_{G}$ the population of the $\ket{G}$ state and $N_{\bar{T}}$ the population of the $\ket{\bar{T}}$ state. We assume that both ground states are populated with equal probability immediately after the first carrier injection pulse. We then numerically solve the rate equation system to obtain the dynamical evolution of the state populations for the duration of the "Control" pulse (7 ns) as a function of the "Control" pulse power. We calculate the polarization contrast (and hence the phase shift) using the ratio $\Xi = \frac{N^i_{\bar{G}}-N^f_{\bar{G}}}{N^i_{\bar{G}}+N^f_{\bar{G}}}$, where $N^i_{\bar{G}}$ and $N^f_{\bar{G}}$ are the populations of the $\ket{\bar{G}}$ state before and after the "Control" pulse respectively. In the absence of the "Control" pulse the experimentally measured phase shift is $78 \pm 4\degree$ (Fig. \ref{Fig3}b). We attribute the non-zero polarization contrast to a certain degree of ellipticity of the intended circularly-polarized ``Target" pulse \cite{androvitsaneas2017efficient}. To account for this, we normalize the calculated ratio $\Xi$ in the absence of the "Control" pulse to the above value.   

In our model we use $\Gamma_1 = 1.2$ GHz and $\Gamma_2 = 1$ GHz. The chosen relaxation rates from the trion states are typical for this kind of QD \cite{PhysRevX.8.021078}. We consider a slightly enhanced rate for the diagonal transition as it is partially coupled to the cavity mode. We assume that $\xi$ accounts for decoherences including spin flip rates and phonon assisted transitions, and is temperature dependent \cite{Anton:17,PhysRevB.80.201311}. It therefore takes the form $\xi = Ae^{\frac{-E_z}{k-BT}}$ \cite{Gerardot.2008}, where $E_z$ is the Zeeman energy, T is the temperature, $k_B$ is the Boltzmann constant and A is a free parameter. We use $E_z$ = 400 $\mu$eV (extracted from Fig. 1d in the main text). The calculated phase shifts as a function of the average photon number per "Control" pulse for different temperatures are shown in Fig. 3b of the main text. For T = 18 K and A = 600 MHz \cite{Climente.2013} we obtain good agreement between the experimental and theoretical values (blue line Fig. 3b of the main text).

\end{document}